\documentclass{ws-procs961x669}            
\usepackage[colorlinks,citecolor=blue, linkcolor=blue]{hyperref}

\begin{document}
\title{Mid-Frequency Gravitational Waves (0.1-10 Hz):\\
	 Sources and Detection Methods\\
Summary of the parallel session GW2 of MG16 Meeting}

\author{Dongfeng Gao$^\#$, Wei-Tou Ni$^*$, Jin Wang$^\$$, Ming-Sheng Zhan$^\dagger$, and Lin Zhou$^\ddagger$}

\address{Wuhan Institute of Physics and Mathematics, APM, Chinese Academy of Sciences, \\
Wuhan 430071, China\\
$^\#$ dfgao@wipm.ac.cn\\
$^*$ wei-tou.ni@wipm.ac.cn\\
$^\$$ wangjin@wipm.ac.cn\\
$^\dagger$ mszhan@wipm.ac.cn\\
$^\ddagger$ lzhou@wipm.ac.cn}

\begin{abstract}
This article summarizes the talks in the session GW2 of {\it the Sixteenth Marcel Grossmann Meeting on Recent Developments in Theoretical and Experimental General Relativity, Gravitation, and Relativistic Field Theories}, 5-10 July, 2021, on {\it Mid-frequency (0.1-10 Hz) gravitational waves: Sources and detection methods} with a review on strain power spectral density amplitude of various mid-frequency gravitational wave projects/concepts and with extended summaries on the progress of ZAIGA project and on the conceptual study of AMIGO.
\end{abstract}


\bodymatter

\section{Introduction }

The mid-frequency GW (Gravitational Wave) band (0.1-10 Hz) between the LIGO-Virgo-KAGRA detection band and LISA-TAIJI/TIANQIN detection band is rich in GW sources. In addition to the intermediate BH (Black Hole) Binary coalescence, it can also come from the inspiral phase of stellar-mass coalescence and from compact binaries falling into intermediate BHs. Detecting mid-frequency GWs enables us to study the compact object population, to test general relativity and beyond-the Standard-Model theories, to explore the stochastic GW background and so on. In addition to DECIGO and BBO, the detection proposals under study include AEDGE, AIGSO, AION, AMIGO, DO, ELGAR, GLOC, INO, LGWA, MAGIS, MIGA, SOGRO, TIAGO, TOBA, ZAIGA, etc. Great advances have accumulated since MG15.

In the GW2 parallel session on 7 July 2021, there were ten talks \cite{ni2021,fier2021,yu2021,jani2021,harms2021,badaracco2021,ando2021,gao2021,aion2021,decigo2021}. First, an outlook of the mid-frequency GW detection and AMIGO were reviewed \cite{ni2021} followed by three talks on the mid-frequency astrophysical/cosmological sources and their propagation through the inhomogeneous universe for the GW observations from the Earth, the Moon and the Space \cite{fier2021,yu2021,jani2021}. The next two talks were on the Lunar Gravitational-Wave Antenna (LGWA) and its sensor development \cite{harms2021,badaracco2021}. The last four talks were on Earth-based detectors TOBA \cite{ando2021}, ZAIGA \cite{gao2021} and AION \cite{aion2021}, and space-borne detectors AEDGE \cite{aion2021}, DECIGO \cite{decigo2021} and B-DECIGO \cite{decigo2021}.

Although there have been activities across all GW frequency spectrum \cite{ni2017}, current activities on GW detection are largely in the high-frequency band (10 Hz-10 kHz) of LIGO-Virgo-KAGRA and low-frequency band (0.1 mHz-100 mHz) of LISA-TAIJI/TIANQIN together with PTAs (300 pHz-100 nHz) and CMB polarization observations (1 aHz-10 fHz). Nevertheless, activities in the mid-frequency GW band have increased steadily since last decade. In the 2030s, ET and Cosmic Explorer will be deployed in the high-frequency band; LISA and TAIJI/TIANQIN will be deployed in the mHz low-frequency band. To complete the map, some mid-frequency GW detectors are desired to be deployed also. Table 1 lists the strain power spectral density (psd) amplitude at 0.1 Hz, 1 Hz and 10 Hz for various mid-frequency GW detectors together with the strain psd amplitude for ET at 1 Hz \& 10 Hz and Cosmic Explorer at 10 Hz, and the strain psd amplitude for LISA at 0.1 Hz and for \& TAIJI \& TIANQIN at 0.1 Hz, 1 Hz and 10 Hz.

\begin{table}
	\tbl{Strain power spectral density (psd) amplitude at 0.1 Hz, 1 Hz and 10 Hz for various mid-frequency GW detectors. In the first column, [E] means Earth-based, [M] means Moon-based, and [S] means space-borne. All units are in ${\rm Hz^{-1/2}}$.}	
	{\begin{tabular}{|c|c|c|c|c|} 
			\hline
			\vspace{2pt}
			GW detector & Method& Strain psd&Strain psd	&Strain psd	\\   
			or&  of   &amplitude&amplitude&amplitude\\
			detector concept& GW detection  &at 0.1 Hz&at 1 Hz &at 10 Hz\\  \hline 
			&Micheson Interferometry& & &\\ 
			ET \cite{et}[E]	& with&n.a.&$1\times 10^{-21}$ &$1\times 10^{-24}$\\ 
			&Fabry-Perot Cavities & & &\\	\hline 	
			&Micheson Interferometry& & &\\ 
			CE \cite{ce}[E]	& with&n.a.& n.a. &$6\times 10^{-25}$\\ 
			&Fabry-Perot Cavities & & &\\	\hline 		
			LISA \cite{lisa}[S]&TDI interferometry&$4\times 10^{-20}$ &$4\times 10^{-19}$ &n.a. \\	\hline 		
			TAIJI \cite{taiji}[S]&TDI interferometry&$3.3\times 10^{-20}$ &$3.3\times 10^{-19}$ &$3.3\times 10^{-18}$ \\ \hline 		
			TIANQIN \cite{tianqin}[S]&TDI interferometry&$1.5\times 10^{-20}$ &$3.5\times 10^{-20}$ &$4\times 10^{-19}$ \\	\hline 
			aSOGRO \cite{asogro}[E]	&SQUID capacitance-&$5\times 10^{-20}$& $3\times 10^{-21}$&$1\times 10^{-21}$\\ 
			&bridge transducer& & &\\	\hline 
			TOBA \cite{toba}[E]&Torsion Bar&$1\times 10^{-19}$ &$8\times 10^{-20}$ &$5\times 10^{-20}$ \\	\hline 
			ELGAR/MIGA \cite{miga}[E]	&Laser-Linked&$3\times 10^{-19}$& $5\times 10^{-22}$&$5\times 10^{-22}$\\ 
			&Atom Interferometry& & &\\	\hline
			MAGIS-1 km \cite{magis}[E]	&Laser-Linked&n.a.& $1\times 10^{-21}$&n.a.\\ 
			&Atom Interferometry& & &\\	\hline
			AION-1 km \cite{aion}[E]	&Laser-Linked&n.a.& $1\times 10^{-21}$&n.a.\\ 
			&Atom Interferometry& & &\\	\hline
			ZAIGA \cite{zaiga}[E]	&Laser-Linked&$1.4\times 10^{-18}$& $2.4\times 10^{-21}$&$1.2\times 10^{-21}$\\ 
			&Atom Interferometry& & &\\	\hline
			LGWA \cite{lgwa}[M]	&Inertial sensors with station&$1\times 10^{-20}$& $1\times 10^{-20}$&n.a.\\ 
			&location measurements& & &\\	\hline
			GLOC \cite{gloc}[M]	&Micheson Interferometry with&n.a.& $1\times 10^{-23}$&$1\times 10^{-24}$\\ 
			&with Fabry-Perot Cavities& & &\\	\hline
			AEDGE \cite{aedge}[S]	&Laser-Linked&$3.5\times 10^{-23}$& $3\times 10^{-23}$&n.a.\\ 
			&Atom Interferometry& & &\\	\hline
			AIGSO \cite{aigso}[S] &Atom Interferometry&$7\times 10^{-19}$ &$1\times 10^{-20}$ &$3\times 10^{-22}$ \\	\hline 
			INO-d \cite{ino}[S] &Clocks with laser link&$1-2\times 10^{-18}$ &$1-2\times 10^{-18}$ &$1-2\times 10^{-18}$ \\	\hline 
			b-AMIGO \cite{amigo}[S] &Michelson X0/X TDI&$5.5\times 10^{-21}$ &$3\times 10^{-21}$ &$6\times 10^{-21}$ \\	\hline 
			AMIGO \cite{amigo}[S] &Michelson X0/X TDI&$4.3\times 10^{-21}$ &$1\times 10^{-21}$ &$2.3\times 10^{-21}$ \\	\hline 
			e-AMIGO \cite{amigo}[S] &Michelson X0/X TDI&$4.3\times 10^{-21}$ &$5\times 10^{-22}$ &$8.5\times 10^{-22}$ \\	\hline 
			TIANGO \cite{tiango}[S] &TDI Interferometry&$1\times 10^{-21}$ &$4\times 10^{-22}$ &$4\times 10^{-22}$ \\	\hline 
			DO-Conservative \cite{do}[S] &TDI Interferometry&$1.5\times 10^{-22}$ &$1.3\times 10^{-22}$ &$3\times 10^{-22}$ \\	\hline 
			DO-Optimal \cite{do}[S] &TDI Interferometry&$4\times 10^{-23}$ &$2\times 10^{-23}$ &$4\times 10^{-23}$ \\	\hline 
			B-DECIGO \cite{decigo2021}[S] &Fabry-Perot arm cavities&$1.5\times 10^{-22}$ &$5\times 10^{-23}$ &$8\times 10^{-23}$ \\	\hline 
			DECIGO 1 cluster \cite{decigo2021}[S] &Fabry-Perot arm cavities&$1.5\times 10^{-23}$ &$5\times 10^{-24}$ &$8\times 10^{-23}$ \\	\hline 
			DECIGO 3-year &Fabry-Perot arm cavities&$3\times 10^{-25}$ &$7\times 10^{-26}$ &$3\times 10^{-25}$ \\	
			correlation \cite{decigo2021}[S]& & & &\\	\hline
			
	\end{tabular} }
	
\end{table}

Harms {\it et al.} \cite{harms2013} have discussed ground-based GW detector concepts to extend the present ground-based interferometers’ detection spectral range to middle-frequency band 0.1–10 Hz. They find that although there are no fundamental limits to the detector sensitivity in this band, the technical limits for the Newtonian-noise cancellation from infrasound and seismic surface fields are challenging to overcome \cite{harms2013,harms2015}. Newtonian noise needs to be measured and subtracted. ET and Cosmic Explorer need to deal with this issue to certain extent and their sensitivity extends only down to 1 Hz and 5 Hz amplitudes respectively. ET is listed at 10 Hz and 1 Hz and Cosmic Explorer is listed at 10 Hz in Table 1 for comparison. On the lower frequency side, LISA, TAIJI and TIANQIN are listed for comparison. For space missions, local gravity noise (Newtonian noise needs to be compensated, and/or measured and subtracted. Due to the success of LISA Pathfinder mission \cite{lisa2016a,lisa2016b}, this issue is largely solved.

The goal on the strain psd noise of ET at 1 Hz is $\sim 1 \times 10^{-21} {\rm Hz^{-1/2}}$, while that of LISA at 0.1 Hz is $4 \times 10^{-20} {\rm Hz^{-1/2}}$. Any new mid-frequency proposals aiming at a starting time after ET and LISA turn-on can keep these sensitivities in mind for their designs.

In Section 2, we discuss and review sources and scientific goals for mid-frequency GW detection. In Section 3, we discuss Earth-based detectors and summarize the TOBA, AION and ZAIGA talks. In Section 4, we discuss Moon-based detectors and summarize the LGWA and GLOC talks. In Section 5, we discuss space-borne GW detectors and summarize the AEDGE, AMIGO, DECIGO and B-DECIGO talks.

\section{Sources and Scientific Goals} 

The general scientific goals of mid-frequency GW detectors are:

(i) to bridge the spectral gap between high-frequency and first-generation low-frequency GW sensitivities for detecting intermediate mass BH coalescence;

(ii) to detect inspiral phase and predict time of stellar-mass binary black hole coalescence together with neutron star coalescence or neutron star-black hole coalescence for ground interferometers, e.g., the inspiral GWs from sources like GW150914; \cite{gw2015}

(iii) to detect compact binary inspirals for studying stellar evolution and galactic population;

(iv) to detect middle frequency GWs from compact binaries falling into intermediate mass BHs for multi-band observation with ground-based GW detectors; \cite{chen2018}

(v) to study the compact object population and to estimate/explore the astrophysical
stochastic background from stellar object inspirals;

(vi) to determine Supermassive Black Hole properties using GW radiation from
surrounding stellar-mass BH binaries; \cite{yu2021,chen2021}

(vii) to test general relativity and beyond-the-standard-model theories.

In subsection 2.1, the goal (vi) will be discussed. General treatments on astrophysical GW sources and associated goals for detection in the middle-frequency GW bands can be found in Ref.’s [35, 36, 38, 39]. In subsection 2.2, Generation, propagation and detection of GWs in inhomogeneous universe will be discussed. In section 3-5, specific scientific goals for individual detector will be stated.

\subsection{Determination of Supermassive Black Hole Properties using GW radiation from surrounding stellar-mass BH binaries\cite{yu2021,chen2021}}

With space-borne gravitational-wave observatories, one may use stellar-mass BH binary as a signal carrier to probe modulations induced by a central supermassive BH (SMBH) to further place constraints on the SMBH's properties. As an example, Yu and Chen \cite{yu2021,chen2018} showed that the de Sitter precession of the inner stellar-mass binary's orbital angular momentum around the angular momentum of the outer orbit would be detectable when the precession period is comparable to the duration of observation, typically a few years. The precession could be combined with the Doppler shift arising from the outer orbital motion to determine the mass of the SMBH and the outer orbital separation with percent-level accuracy. Joint detection with Earth-based detector would extend the detection threshold to a precession period to $\sim$100 yr.

\subsection{Generation, propagation and detection of GWs in inhomogeneous universe}
In the talk, Fier {\it et al.}\cite{fier2021,fier2021b} presented their recent studies on GWs produced by remote compact astrophysical sources. To describe such GWs properly, Fier {\it et al.} introduced three scales, the typical wavelength of GWs, the scale of the cosmological perturbations, and the size of the observable universe. For GWs to be detected by the current and foreseeable detectors, they can be well approximated as high-frequency GWs. The spatial, traceless, and Lorentz gauge conditions can be imposed simultaneously to simplify the field equations, even when the background is not vacuum, as long as the high-frequency GW approximation is valid. Fier {\it et al.} developed the geometrical optics approximation to such GWs, and calculated the gravitational integrated Sachs-Wolfe effects due to the presence of the cosmological scalar and tensor perturbations to read out explicitly the dependence of the amplitude, phase and luminosity distance of the GWs.

\section{Earth-based detectors – TOBA, AION and ZAIGA}

Earth is noisy for strain measurement. From 10 Hz to 10 kHz, vibration noise needs to be suppressed. This issue is basically solved by active and passive vibration isolation after decades of research and development. In the mid-frequency band of GW detection on Earth, the Newtonian noise (the gravity gradient noise) cannot be suppressed. It needs to be measured and subtracted. TOBA uses torsion bars, and AION and ZAIGA use atom interferometry to measure the Newtonian noise and subtract it.

\subsection{TOBA}
TOBA (TOrsion-Bar Antenna) is a mid.-frequency gravitational-wave antenna. It is formed by two bar-shape test masses, each suspended as a torsion pendulum. Tidal effect originated by incoming gravitational wave will be detected as differential angular motion of these two bars. The fundamental sensitivity is $10^{-19} {\rm Hz^{-1/2}}$ at 0.1 Hz frequency band, assuming 10-m scale cryogenic detector at 4 K with the rotations of the bars measured with Fabry–Perot cavities at the end of the bars.\cite{ando2021,toba} Though this sensitivity is not comparable with space antennae, it is sufficient to observe some intermediate-mass black-hole inspirals in our universe. Also, operation on ground is advantageous in development time and cost. The Phase-III prototype is to complete the demonstration of noise reduction reaching a sensitivity of about $10^{-15} {\rm Hz^{-1/2}}$ at 0.1 Hz. Measuring the terrestrial gravity fluctuation with a sensitivity of about or below $10^{-15} {\rm Hz^{-1/2}}$ at 0.1 Hz is useful for two geophysical purpose – earthquake early warning and Newtonian noise reduction for GW detection.

\subsection{AION}
AION (Atom Interferometer Observatory and Network) is a proposed experimental programme using cold strontium atoms to search for ultra-light dark matter, to explore gravitational waves in the mid-frequency, and to probe other frontiers in fundamental physics. AION has 3 stages of development, AION-10 m, AION-100 m and AION-1 km in parallel to MAGIS-10 m, MAGIS-100 m and MAGIS-1 km. AION would share many technical features with the MAGIS experimental programme, and synergies would flow from operating AION in a network with MAGIS, as well as with other atom interferometer experiments such as MIGA, ZAIGA and ELGAR.\cite{aion2021,aion}

\subsection{ZAIGA}
ZAIGA (the Zhaoshan long-baseline Atom Interferometer Gravitation Antenna) is a proposed underground long-baseline atom interferometer (AI) facility (see Fig. 1), aiming for experimental research on gravitation and related problems.\cite{zaiga} It is located in the mountain named Zhaoshan which is about 80 km southeast to Wuhan city. ZAIGA will consist of gravitational wave detection (ZAIGA-GW), dark matter detection (ZAIGA-DM), high-precision test of the equivalence principle (ZAIGA-EP), clock-based gravitational red-shift measurement (ZAIGA-CE-R), rotation measurement and gravitomagnetic effect (ZAIGA-RM), and geological and geophysical measurement (ZAIGA-GG). The first stage of the project (2021-2025) will include building a vertical 300-meter long tunnel to interrogate three 10-meter atom interferometers and atomic clocks (aiming for ZAIGA-EP, ZAIGA-DM and ZAIGA-CE-R), and building a horizontal 1.5-km long tunnel (aiming for ZAIGA-RM and ZAIGA-GG). Depending on the performance in the first stage, the second stage of the project will construct an array of 1-km arm-length horizontal triangular tunnels and be dedicated to ZAIGA-GW.

\begin{figure}[htbp]
	\centering
	\includegraphics[width=0.9\textwidth]{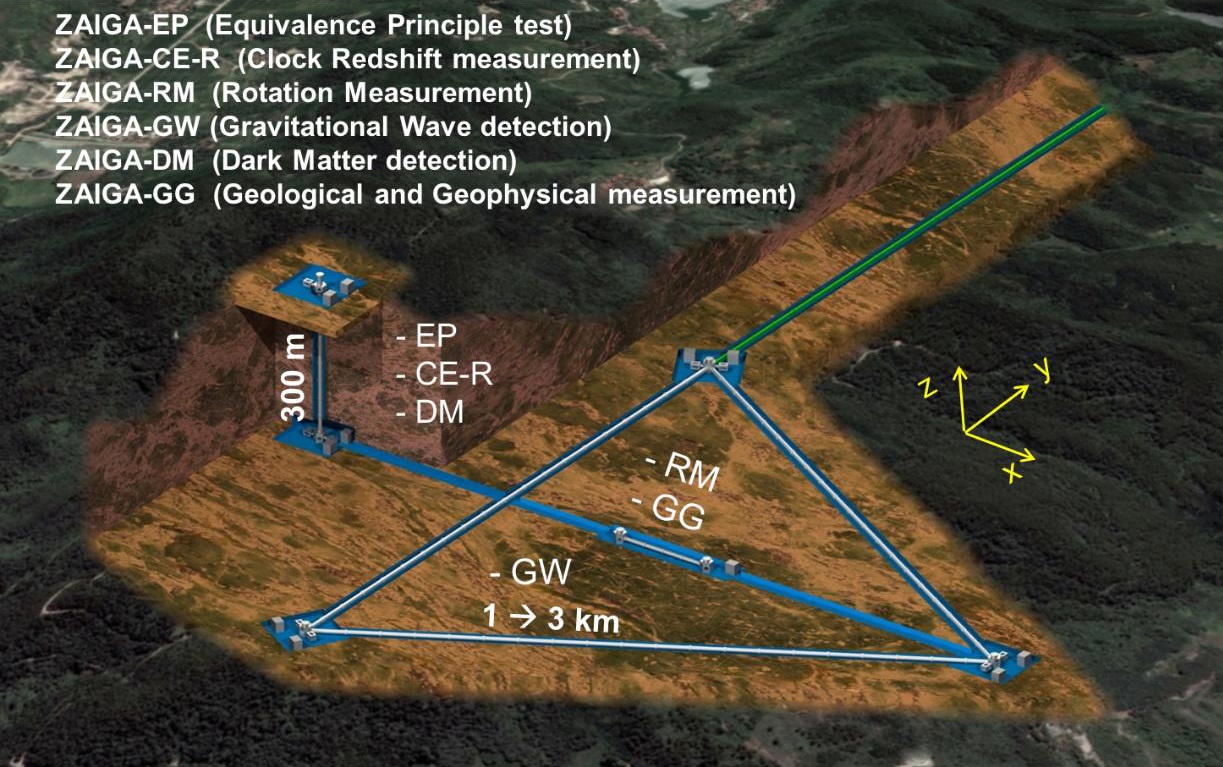}
	\caption{Schematic diagram of ZAIGA design}    \label{ZAIGA} 
\end{figure}

The site exploration started immediately after the ZAIGA plan was put forward. Relevant work for site exploration keeps on going now. The following progress on the technical and theoretical aspects of ZAIGA has also been made.

In the past three years, with our 10-m dual-species atom interferometer, we carried out a joint mass-energy test of the equivalence principle (EP), by using rubidium atoms with specified mass and internal energy.\cite{zhou2021} By extending the four-wave double-diffraction Raman transition method (4WDR) to $^{85}$Rb and $^{87}$Rb atoms with different angular momenta, we can measure their differential gravitational acceleration. The Eötvös parameter, $\eta$, of the four paired combinations ($^{85}Rb|F=2\rangle-^{87}Rb|F=1\rangle$, $^{85}Rb|F=2\rangle-^{87}Rb|F=2\rangle$, $^{85}Rb|F=3\rangle-^{87}Rb|F=1\rangle$ and $^{85}Rb|F=3\rangle-^{87}Rb|F=2\rangle$) were measured to be $\eta_1=(1.5\pm 3.2)\times 10^{-10}$, $\eta_2=(-0.6\pm 3.7)\times 10^{-10}$, $\eta_3=(-2.5\pm 4.1)\times 10^{-10}$ and $\eta_4=(-2.7\pm 3.6)\times 10^{-10}$, respectively. The violation parameters of mass and internal energy are constrained to $\eta_0=(-0.8\pm 1.4)\times 10^{-10}$ and $\eta_E=(0.0\pm 0.4)\times 10^{-10}$. This work opens a door for joint tests of two attributes beyond the traditional pure mass or energy tests of EP with quantum systems. It also provides us a firm technical foundation for our proposed ZAIGA-EP experiment. For the rotation measurement, Yao {\it et al.} in our group has improved the precision of the measurement by one-order of magnitude recently and keep on advancing the precision and accuracy.\cite{yao2021}

Recently, a theoretical study on the possibility of detecting the ultralight scalar dark matter (DM) with ZAIGA was completed.\cite{zhao2021} Starting with a popular scalar DM model, where the DM field is assumed to linearly couple to the standard model fields through five coupling parameters, the DM candidate can be detected by determining whether the five coupling parameters are zero or not. The solution to the DM field contains a background oscillation term and a local exponential fluctuation term. The DM signals in ZAIGA has been calculated. For a pair of AIs vertically separated by 300 meters, the DM-induced differential phase consists of three contributions, coming from the DM-induced changes in atomic internal energy levels, atomic masses and the gravitational acceleration. For a pair of AIs horizontally separated by several kilometers, the signal comes only from the DM-induced changes in atomic internal energy levels. The constraints on five DM coupling parameters for ZAIGA-DM have been estimated. It turns out that the proposed constraints would complement the MICROSCOPE space experiment\cite{microscope2020} and would be several orders of magnitude better in higher-mass parametric regions. It also clearly shows the advantages and disadvantages of the two configurations of ZAIGA-DM. For the vertical configuration, the advantage is that all the five DM coupling parameters can be constrained, and the disadvantage is that the arm-length is not easy to extend. For the horizontal configuration, the advantage is that the arm-length is relatively easy to extend, and the disadvantage is that only two DM parameters can be constrained. For other type of DM candidates, the relevant studies are in progress.

\section{Moon-based detectors – LGWA and GLOC}
Moon is much quieter seismologically and does not have an atmosphere. This makes Moon itself a possible good detector by monitoring its surface motion excited by GWs. This also makes an interferometer GW detector based on the Moon possible.

\subsection{Lunar Gravitational-Wave Antenna\cite{harms2021,lgwa}}
One of the first concepts proposed for the detection of gravitational waves by Joseph Weber is to monitor the quadrupolar vibrations of elastic bodies excited by GWs. At laboratory scale, these experiments became known as resonant-bar detectors, which form an important part of the history of GW detection. Due to the dimensions of these bars, the targeted signal frequencies were in the kHz range. It was also Weber who suggested to monitor vibrations of Earth and Moon to search for gravitational waves in the mHz band. His Lunar Surface Gravimeter was deployed on the Moon in 1972 by the Apollo 17 crew. A design error made it impossible to carry out the intended search for GWs, but the idea remains intriguing. Harms {\it et al.}\cite{harms2021,lgwa} have proposed a new concept, the Lunar Gravitational-Wave Antenna (LGWA), based on Weber’s idea. LGWA would have a rich GW and multi-messenger science case with galactic binaries and more massive black-hole binaries. It would also serve as a high-precision geophysical station shedding light on the interior structure of the Moon, the mechanisms of moonquakes, and the Moon's formation history. The key component is a next-generation, high-sensitivity seismometer to be deployed on the Moon. For its most sensitive realization, LGWA would have to be deployed in a permanent shadow near the south or north pole of the Moon to benefit from the natural cryogenic environment. This would improve the sensitivity of the seismometer and also provide a lower-noise environment due to the absence of thermally induced seismic events that were observed by the Apollo seismometers. Powering of the seismic stations and data transfer pose additional challenges for such a deployment.

\subsubsection{Inertial sensors for the LGWA \cite{badaracco2021}}
The core of LGWA will be composed of an array of high-end seismic sensors: CSIS (Cryogenic Superconducting Inertial Sensor) on the Moon’s surface. A cryogenic environment will be used in combination with superconducting materials to open up pathways to low-loss actuators and sensor mechanics.

CSIS revolutionizes the (cryogenic) inertial sensor field by obtaining a displacement sensitivity at 0.5 Hz of 3 orders of magnitude better than current state-of-art. It will allow LGWA to be sensitive below 1 Hz, down to 1 mHz. It will also be employed in the forthcoming Einstein Telescope (ET) - a third-generation gravitational-wave detector that will make use of cryogenic technologies and will have an enhanced sensitivity below 10 Hz. Moreover, CSIS seismic data could also be employed to get new insights about the Moon’s interior… and the selenphysics (the Moon’s geophysics). A technical design for CSIS with both interferometric readout and Rasnik readout is presented at the meeting.

\subsection{GLOC (Gravitational-Wave Lunar Observatory for Cosmology)\cite{jani2021,gloc}}
Taking advantage of the vacuum environment and the low seismic activities of the Moon, GLOC proposes to build a full laser-interferometric triangle-shape GW detector with 40 km arm length to explore fundamental cosmology and multi-band, multi-messenger astrophysics from the Moon. The speaker focused on goals that are unique to GLOC, and compare the detection landscape of the elusive intermediate-mass black holes between GLOC and other prominent space-based missions, 3G detectors and deci-Hertz concepts.

\section{Space-borne GW detectors}
The space environments are much quieter in terms of the gravity gradient noise consideration as compared to the Earth and Moon environments. It is so quiet that the major gravity gradient noise comes from local environments inside the spacecraft. In space, the GW strains propagating to the detector are measured above the geodetic motion or the desired controlled motion/position of the spacecraft. For spacecraft to follow geodesics, one needs to perform drag-free control. For spacecraft to follow desired motion/position over the geodetic motion, one needs to actuate acceleration on the geodetic motion in addition to basic drag-free control. LISA Pathfinder launched on December 3, 2015 has completely met the LISA drag-free noise requirement and has successfully demonstrated the first generation drag-free technology requirement for space detection of GWs.\cite{lisa2016a,lisa2016b}

Many space GW detection proposals need to use constant/equal arm configurations. They are AEDGE,\cite{aion2021,aedge} AIGSO,\cite{aigso,aigso2020} DECIGO/B-DECIGO,\cite{decigo2021} etc. AIGSO has 10 km arm length, the shortest arm length among these mission proposals. The calculated actuation acceleration needed to maintain such orbits for AIGSO is around 10 pm s$^{-2}$.\cite{aigso2020,wang2020}

In the mission of LISA Pathfinder, different levels of force and torque authority were implemented, from the nominal configuration with x-force authority (on the sensitive line-of-sight axis) of 1100 pm s$^{-2}$ to the URLA configuration levels, with x-force authority of 26 pm s$^{-2}$.\cite{lisa2019} The published LPF differential acceleration noise floor is established by measurements in this configuration. Specifically, LISA Pathfinder demonstrated that when a constant out of the loop force with amplitude of 11.2 pN was applied to the sensitive axis of TM1 (Test Mass 1) for reducing the gravitational imbalance between the TMs, this force does not introduce significant noise or calibration errors.\cite{lisa2019} Basically, the accelerometer part of the constant-arm technology is already demonstrated by LISA Pathfinder for AIGSO.

B-DECIGO has a nominal arm length of 100 km, DECIGO 1,000 km, and AMIGO 10,000 km. The actuation accelerations needed are respectively 10, 100, and 1000 times more than AIGSO. While the actuation accelerations needed for constant arm implementation of b-DECIGO and DECIGO is still basically in the LISA Pathfinder nominal configuration range, the actuation accelerations for constant arm AMIGO is one order larger. On what noise level could the actuation accelerations be done is an issue that needs to be studied and demonstrated carefully for AMIGO. A suggestion is to use an additional test mass (i.e. a pair) to alternate with the original one. \cite{wang2020,ni2020}

GW space missions do not usually require large communication capacity. The deployment to 2-4 degrees behind or ahead of the Earth orbit is no more complicated as to L1 or L2 Sun-Earth Lagrange points. We hereby provide {\it an example for taking less than a week to reach the (pre)-science orbits}. A last-stage launch from 300 km LEO (Low Earth Orbit) to an eccentric Hohmann orbit with apogee 262931 km (The period of this Hohmann transfer orbit is about 6 days.) (Fig. 2). It takes 3 days (half an elliptic orbit) for this transfer from perigee to apogee (Fig. 2. (b)). This apogee can be designed to be the closest encounter point with respect to Earth of the center of mass of the 3 S/C traced back in time geodetically of the 2-degree-behind-the-Earth formation. From here, a $\Delta {\rm v_1}$'s of about 1.6 km/s are needed to allow the 3 S/C to enter their respective (pre-) science orbits (Fig. 2. (c)) when the calibration, commissioning and various science operations can be started. For example, it is the AMIGO-S-2-4deg orbit for AMIGO and the time epoch to reach the 2-degree-behind-the-Earth point is 2030-Jan-1st 12:00:00.\cite{ni2020} (AMIGO-Earth-like solar orbits with the AMIGO formation varying between 2 and 4 degrees behind the Earth orbit starting at epoch JD2462503.0 (2030-Jan-1st 12:00:00) in J2000 mean-equatorial solar-system-barycentric coordinate system.) There are similar orbit configurations for other time epochs and for ahead of the Earth situations. For ahead of the Earth situations, the apogees of the Hohmann orbits need to be on the other side of the Earth.

In the following, we summarize AEDGE,\cite{aion2021} AMIGO,\cite{ni2021} and DECIGO and B-DECIGO\cite{decigo2021} in our session.

\subsection{AEDGE \cite{aion2021,aedge}}
Atomic Experiment for Dark Matter and Gravity Exploration (AEDGE) is a concept for a space experiment using cold atoms to search for ultra-light dark matter, and to detect gravitational waves in the mid-frequency band. It will also complement other planned searches for dark matter, and exploit synergies with other gravitational wave detectors. Examples are given in the talk of how its gravitational-wave measurements could explore the assembly of super-massive black holes, first-order phase transitions in the early universe and cosmic strings.\cite{aion2021,aedge} AEDGE will be based upon technologies now being developed for terrestrial experiments using cold atoms, and will benefit from the space experience obtained with, e.g., LISA and cold atom experiments in microgravity.

\subsection{AMIGO}

In 2017, a middle-frequency GW mission AMIGO (Astrodynamical Middle-frequency Interferometric GW Observatory) with arm length 10000 km was proposed to have significant sensitivity in this frequency band 0.1 Hz-10 Hz to bridge the gap between the ground detection sensitivity and the LISA sensitivity and yet to be a first-generation candidate for space GW missions.\cite{amigo2018} The basic mission concept is in Ref. [51]. Stellar-size black hole inspirals \cite{gw2015} such as the ones like GW150914 will be sure sources of AMIGO. The AMIGO observation of the inspiral phase of these sources will enable us to predict the coalescence time precisely for the ground multi-messenger observations. 

\begin{figure}[htbp]
	\centering
	\includegraphics[width=0.9\textwidth]{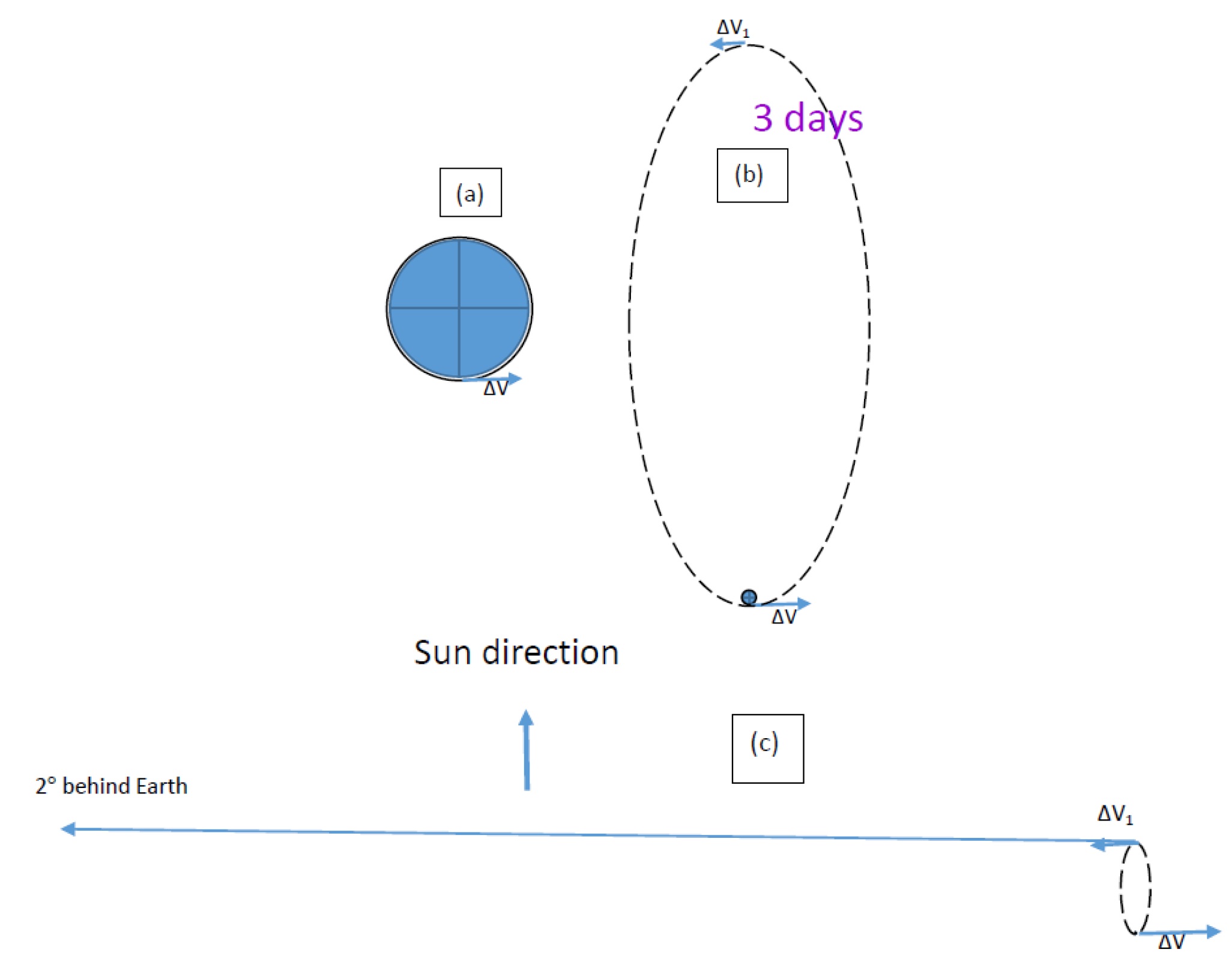}
  \caption{An option for taking less than a week from launch to reaching the (pre-)science orbit for AMIGO-S-2-4 deg. (a) Launch to 300 km LEO (Low Earth Orbit) parking orbit. (b) A last-stage launch from 300 km LEO to an eccentric Hohmann orbit with apogee 262931 km (The period of this Hohmann transfer orbit is about 6 days.); It takes 3 days (half an elliptic orbit) for this transfer from perigee to apogee. (c) This apogee is designed to be the closest encounter point with respect to Earth of the center of mass of the 3 S/C traced back in time geodetically of the 2-degree-behind-the-Earth AMIGO formation]. From here, $\Delta {\rm v_1}$'s of about 1.6 km/s are needed to allow the 3 S/C to enter their respective (pre-)science orbits reaching 2-degree-behind-the-Earth at 2030-Jan-1st 12:00:00.\cite{amigo, wu2017,wu2021}}    \label{AMIGO1} 
\end{figure}

In [48], first studies on possible schemes of implementation of AMIGO were presented. Both the solar orbit and earth orbit options together with deployment strategy were discussed. The first-generation TDIs (Time Delay Interferometry’s) for all orbit options studied were calculated and found that all the heliocentric orbit options satisfy the frequency-noise suppression requirement, but the geocentric orbit options do not satisfy the requirement. From this study, the heliocentric option is preferred. The issue on feasibility of constant equal-arm implementation was studied. For the solar-orbit options, the acceleration to maintain a constant equal-arm formation can be designed to be less than 15 nm/$s^2$ with the thruster requirement in the 15 $\mu$N range.\cite{wang2020,ni2020} AMIGO would be a good place to test and implement the constant equal-arm option. Fuel requirement, thruster noise requirement and test mass acceleration actuation requirement were considered.

In [28], the core noise requirements on position noises and acceleration noises were updated. From these design white position noises and acceleration noises, the GW sensitivities (Fig. 3) were obtained for the first-generation Michelson X TDI configuration of b-AMIGO (baseline AMIGO), AMIGO, and e-AMIGO (enhanced AMIGO). In Fig.3, the sensitivities for the corresponding AMIGO-5’s (50,000 km nominal arm length) are also shown. Steps to implement the AMIGO mission concept were indicated in view of the current technology development.

\begin{figure}[htbp]
	\centering
	\includegraphics[width=0.9\textwidth]{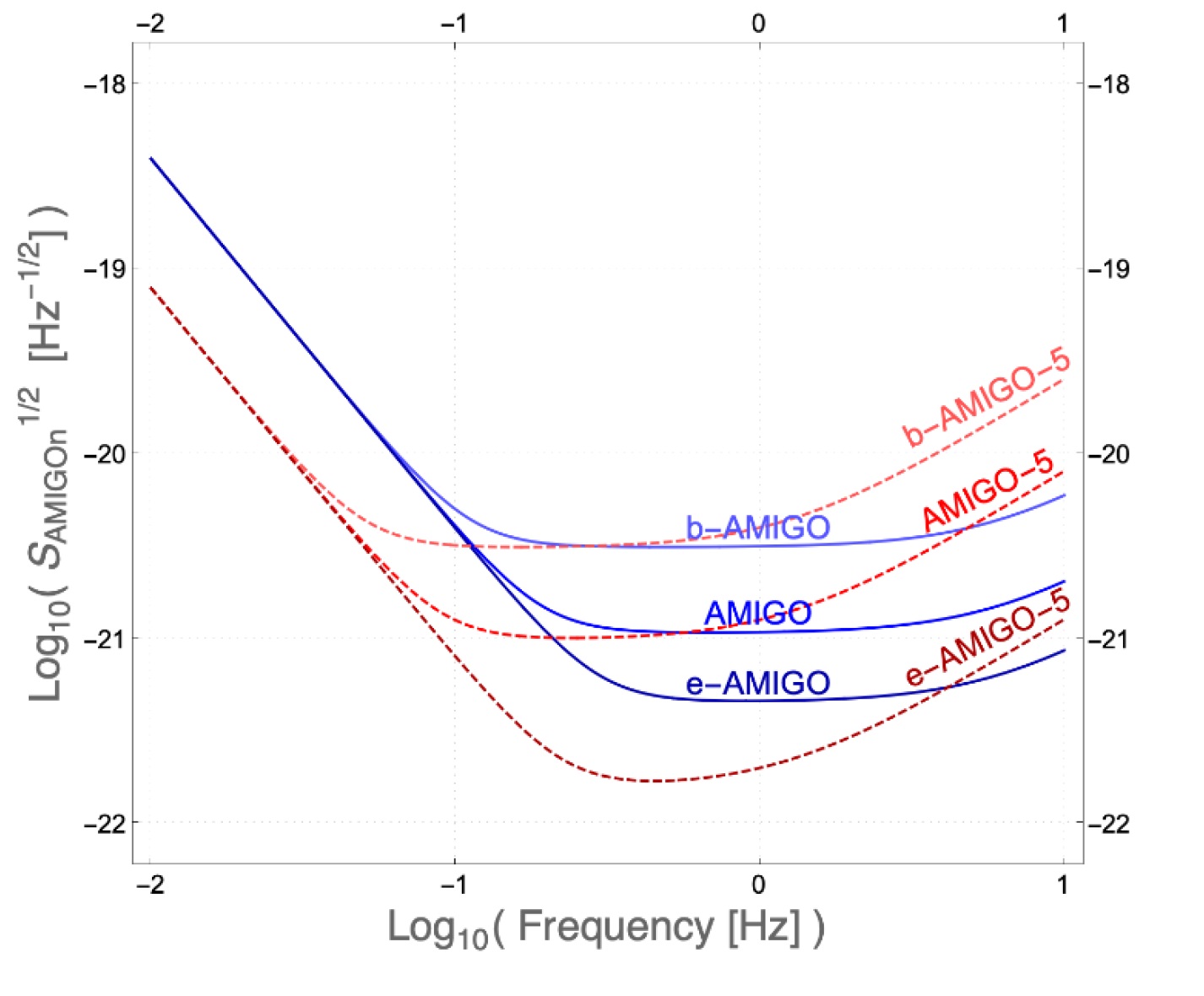}
	\caption{Strain power spectral density (psd) amplitude vs. frequency for various AMIGO-5 proposals as compared to AMIGO proposals. The solid lines are for AMIGO’s (10,000 km nominal arm length); the dashed lines are for AMIGO-5’s (50,000 km nominal arm length).\cite{amigo}}    \label{AMIGO2} 
\end{figure}

\subsection{DECIGO and B-DECIGO \cite{decigo2021}}
DECIGO was conceived in 2001 with the main scientific target to directly measure the acceleration of the Universe. The design has been elaborated with laser Fabry-Perot arm cavities. DECIGO consists of four clusters with each cluster having three differential Fabry-Perot interferometers with three drag-free spacecraft forming an equilateral triangle of nominal arm cavity length 1000 km. Lasers illuminating the cavity have power 10 W and wavelength 515 nm. The current design has main mirrors with diameter 1 m and mass 100 kg forming cavities of finesse 10. The four clusters also form a nearly equilateral triangle in the heliocentric orbits with two clusters in Earth-trailing orbits. The other two clusters are separated from each other and from the near-Earth clusters by 120 degrees in the heliocentric orbits. The DECIGO team recognized that “… analysis based on the observations by the Planck satellite lowered the upper limit of GW to $1\times 10^{-16}$ As a result, the DECIGO sensitivity became no longer good enough to detect the primordial gravitational waves”. Nevertheless, with target sensitivity the best among all the present mid-frequency concepts (Table 1), it may still bring us some clues on the very early universe, e.g. revealing the thermal history after the inflation \cite{cosmo2015,cosmo2008} etc.

The design of B-DECIGO is down-scaled from DECIGO, consisting of one cluster with three spacecraft separated from each other by 100 km. The arm cavity mirrors have diameter of 0.3 m with mass of 30 kg, and cavity finesse of 100. The laser wavelength is 515 nm and laser power 1 W. With target strain psd amplitude about one order higher than DECIGO, it serves as a pathfinder for DECIGO, and with the noise sensitivity well enough, it is also a good probe to the many mid-frequency GW sources discussed in section 2. DECIGO team aim to launch B-DECIGO at the earliest in 2032.

\section{Discussion and Outlook}
Middle frequency band (0.1-10 Hz) is important for GW detection to bridge the gap between high-frequency GW detection and low-frequency GW detection for various scientific goals. We have seen various proposals. It is hopeful to have some mid-frequency GW detectors in the 2030s to do multiband astronomy and multi-messenger astronomy.

\section*{Acknowledgements}
This work was supported by the Strategic Priority Research Program of the Chinese Academy of Sciences under grant No. XDB21010100, and by the National Key Research and Development Program of China under Grant No. 2016YFA0302002.




\end{document}